\documentclass[preprint,showpacs,preprintnumbers,amsmath,amssymb]{revtex4}

\usepackage{amsfonts}
\usepackage{graphicx}
\usepackage{dcolumn}
\usepackage{axodraw}

\providecommand{\eps}{\epsilon}
\providecommand{\epsl}{\epsilon^{\prime}}

\begin{document}
\renewcommand{\baselinestretch}{0.95}


\title{ Soft CP violation in K-meson systems}

\author{J. C. Montero}
\email{montero@ift.unesp.br}
\author{C. C. Nishi}
\email{ccnishi@ift.unesp.br}
\author{V. Pleitez}
\email{vicente@ift.unesp.br}
\affiliation{Instituto de F\'\i sica Te\'orica,
Universidade Estadual Paulista\\
Rua Pamplona, 145\\
01405-900-- S\~ao Paulo\\
Brazil}
\author{O. Ravinez}
\email{opereyra@uni.edu.pe}
\affiliation{
Universidad Nacional de Ingenieria UNI,
Facultad de Ciencias\\
Avenida Tupac  Amaru S/N apartado 31139\\
Lima, Peru}
\author{M. C. Rodriguez}
\email{mcrodriguez@fisica.furg.br}

\affiliation{
Funda\c c\~ao Universidade Federal do Rio Grande/FURG,
Departamento de F\'\i sica\\
Av. It\'alia, km 8, Campus Carreiros \\
96201-900, Rio Grande, RS\\
Brazil }
\date{\today}

\begin{abstract}
We consider a model with soft CP violation which accommodate the CP violation in
the neutral kaons even if we assume that the Cabibbo-Kobayashi-Maskawa mixing
matrix is real and the sources of CP violation are three complex vacuum
expectation values and a trilinear coupling in the scalar potential. We show
that for some reasonable values of the masses and other parameters the model
allows to explain all the observed CP violation processes in the
$K^0$-$\bar{K}^0$ system.  
\end{abstract}

\maketitle

\section{Introduction}
\label{sec:intro}

Until some time ago, the only physical system in which
the violation of the CP symmetry was observed was the neutral kaon
system~\cite{kaons}. Besides, only the indirect CP violation described by
the $\epsilon$  parameter was measured in that system.
Only recently, clear evidence for direct CP violation parametrized  by the
$\epsilon^\prime$ parameter was observed in laboratory~\cite{direct}.
Moreover, the CP violation in the $B$-mesons system has been, finally,
observed as well~\cite{bottom}. It is in fact very impressive that all of 
these observations are accommodated by the electroweak standard model with 
a complex Cabibbo-Kobayashi-Maskawa mixing matrix~\cite{km,ckmfitter} when 
QCD effects are also included. In the context of that model, the
only way to introduce CP violation is throughout its hard violation due to 
complex Yukawa couplings, which imply a surviving phase in the charged
current coupled to the vector boson $W^{\pm}$ in the quark sector. In the 
neutral kaon system, despite the CKM phase being $O(1)$,  the breakdown of 
that symmetry is naturally small because its effect involves the three 
quark families at the one loop level~\cite{kayser}. This is not the case 
of the $B$ mesons where the three families are involved even at the tree 
level and the CP violating asymmetries are $O(1)$~\cite{soni05}. 

Notwithstanding, if new physics does exist at the TeV scale it may imply new
sources of CP violation. In this context the question if the CKM matrix is
complex becomes nontrivial since at least part of the CP violation may come from
the new physics sector~\cite{botella}. For instance, even in the context of a 
model with $SU(2)_L\otimes U(1)_Y$ gauge symmetry, we may have spontaneous CP
violation through the complex vacuum expectation values (VEVs), this is the case 
of the two Higgs doublets extension of the standard model if we do not 
impose the suppression of flavor changing neutral currents (FCNCs), as in 
Ref.~\cite{tdl}. The CP violation may also 
arise throughout the exchange of charged scalars if there are at least three
doublets and no FCNCs~\cite{sw}. Truly soft CP 
violation may also arise throughout a complex dimensional coupling constant 
in the scalar potential and with no CKM phase~\cite{lavoura}. In fact, all 
these mechanisms can be at  work in multi-Higgs extensions of the standard 
model~\cite{wu}. Hence, in the absence of a general principle, all possible 
sources of $CP$ violation must be considered in a given model. However, 
it is always interesting to see the potentialities of a given source to 
explain by itself all the present experimental data. This is not a trivial 
issue since, for instance, CP violation mediated by Higgs scalars in models 
without flavor changing neutral currents have been almost ruled out even by 
old data~\cite{desphande,sanda,donoghue,wolfen,chang}. 

Among the interesting extensions of the standard model there are the models 
based on the $SU(3)_C\otimes SU(3)_L\otimes U(1)_X$ gauge 
symmetry called 3-3-1 models for short~\cite{331,pt,mpp}. 
These models have shown to be very predictive 
not only because of the relation with the generation problem, some 
representation content of these models allow three and only three families
when the cancellation of anomalies and asymptotic freedom are used;
they also give some insight about the observed value of the weak 
mixing angle~\cite{pl331}. The 3-3-1 models are also 
interesting context in which new theoretical ideas as 
extra dimensions~\cite{extra} and the little Higgs mechanism can be
implemented~\cite{lhiggs}.

In the minimal 3-3-1 model~\cite{331} both 
mechanisms of CP violation, hard~\cite{liung} and spontaneous~\cite{laplata1} 
has been already considered. In this paper we analyze soft CP violation in 
the framework of the 3-3-1 model of Ref.~\cite{pt} in 
which only three triplets are needed for breaking the gauge symmetry 
appropriately and give mass to all fermions. Although it has been 
shown that in this model pure spontaneous CP violation is not
possible~\cite{laplata1}, we can still implement soft CP violation 
if, besides the three scalar VEVs, a trilinear parameter in the scalar 
potential is allowed to be complex. In this case a physical phase survives 
violating the CP symmetry. This mechanism was developed in Ref.~\cite{cp3} 
but there a detailed analysis of the CP observables in both kaons and 
$B$-mesons were not given. Here we will show that all the CP violating 
parameters in the neutral kaon system can be explained through this 
mechanism leaving the case of the $B$-mesons for a forthcoming paper.

The outline of this paper is as follows. In Sec.~\ref{sec:model} we briefly
review the model of Ref.~\cite{cp3} in which we will study a mechanism for soft
CP violation. In Sec.~\ref{sec:cpvkaons} we review the usual parameterization of
the CP violating parameters of the neutral kaon system, $\epsilon$ and
$\epsilon^\prime$, establishing what is in fact that is being
calculated in the context of the present model. 
In Sec.~\ref{sec:epsilonprime} we calculate $\epsilon$,
and in Sec.~\ref{sec:epsilon} we do the same for $\epsilon^\prime$. The possible
values for those  parameters in the context of our model are considered in
Sec.~\ref{sec:fitting}, while our conclusions are in the last section. In the
Appendix A we write some integrals appearing in box and penguin diagrams.  

\section{The model}
\label{sec:model}

Here we are mainly concerned with the doubly charged scalar and its Yukawa
interactions with quarks since this is the only sector in which the soft $CP$
violation arises in this model~\cite{cp3}. The interaction with the
doubly charged vector boson will be considered when needed
(Sec.~\ref{sec:epsilon}). As expected, there is only a doubly charged would be
Goldstone boson, $G^{++}$, and a physical doubly charged scalar, 
$Y^{++}$, defined by
\begin{equation}
\left(
\begin{array}{c}
\rho^{++} \\ \chi^{++}
\end{array}
\right)=\frac{1}{N}\,
\left(\begin{array}{cc}
\vert v_\rho\vert & -\vert v_\chi\vert\,e^{-i\theta_\chi}
\\ \vert v_\chi\vert\,e^{i\theta_\chi}  & \vert v_\rho\vert
\end{array}
\right)
\left(
\begin{array}{c}
G^{++} \\ Y^{++}
\end{array}
\right),
\label{dcf}
\end{equation}
where $N= (\vert v_\rho\vert^2+\vert v_\chi\vert^2)^{1/2}$; the mass
square of the $Y^{++}$ field is given by
\begin{equation}
m^2_{Y^{++}}=\frac A{\sqrt{2}}\left( \frac 1{\vert v_\chi\vert^2}+
\frac1{\vert v_\rho \vert ^2}\right)
 -\frac{a_8}2\left( \vert v_\chi\vert ^2+\vert v_\rho \vert^2\right),
\label{m++}
\end{equation}
where we have defined $A\equiv \mbox{Re}(f v_\eta v_\rho v_\chi)$
with $f$ a complex parameter in the trilinear term $\eta\rho\chi$ of the
scalar potential and $a_8$ is the coupling of the quartic term
$(\chi^\dagger\rho)(\rho^\dagger\chi)$ in the scalar potential. For details and
notation see Ref.~\cite{cp3}.
Notice that since $\vert v_\chi\vert \gg \vert v_\rho\vert$, it is $\rho^{++}$
which is almost $Y^{++}$.

In Ref.~\cite{cp3} it was shown that all $CP$ violation effects arise from
the singly and/or doubly charged scalar-exotic quark interactions.
Notwithstanding, the CP
violation in the singly charged scalar is avoided by assuming the total leptonic 
number L (or B+L, see below) conservation and, in this case, only two phases 
survive after the re-definition of the phases of all fermion fields in the model:
a phase of the trilinear coupling constant $f$ and the phase of a vacuum 
expectation value, say $v_\chi$. Among these phases, actually only one survives 
because of the constraint equation
\begin{equation}
{\rm Im}\left( f v_\chi v_\rho v_\eta \right) =0,
\label{vin4}
\end{equation}
which implies $\theta_\chi=-\theta_f$.

Let us briefly recall the representation content of the model~\cite{cp3}
with a little modification in the notation. In the quark sector we have
$Q_{iL}=(d_i,\,u_i,j_i)^T_L\sim({\bf3},{\bf3}^*,-1/3),\;i=1,2$
$Q_{3L}=(u_3,d_3,J)^T_L\sim({\bf3},{\bf3},2/3)$;
$U_{\alpha R}\sim({\bf3},{\bf1},2/3)$, $D_{\alpha R}\sim({\bf3},{\bf1},-1/3),
\,\alpha=1,2,3$, $j_{iR}\sim({\bf3},{\bf1},-4/3)$ and
$J_R\sim({\bf3},{\bf1},5/3)$, and the Yukawa interactions are
written as:
\begin{eqnarray}
-{\cal L}&=&\sum_{i\alpha}\overline{Q_{iL}}\,(F_{i\alpha}\rho^*U_{\alpha
R}+\tilde{F}_{i\alpha}D_{\alpha R}\eta^*)
+\overline{Q_{3L}}
(\,F_{3\alpha}U_{\alpha R}\eta
 +\tilde{F}_{3\alpha}D_{\alpha R}\rho )
\nonumber \\ &+&
\sum_{im}\lambda_{im}\overline{Q_{iL}}\,j_{mR}\chi^*
+\lambda_3\overline{Q_{3L}}J_R\chi 
+H.c.,
\label{yukawa}
\end{eqnarray}
where all couplings in the matrices $F,\tilde{F}$ and $\lambda$'s are in principle 
complex. Although the fields in Eq.~(\ref{yukawa}) are symmetry eigenstates
we have omitted a particular notation. Here we will assume
that all the Yukawa couplings in Eq.~(\ref{yukawa}) are real in such a way that
we may be able to test to what extension only the phase $\theta_\chi$ can
describe the CP violation parameters in the neutral kaon system, $\epsilon$
and $\epsilon^\prime$. 

In order to diagonalize the mass matrices coming from Eq.~(\ref{yukawa}), we introduce
real and orthogonal left- and right-handed mixing matrices defined as
\begin{equation}
U^\prime_{L(R)}={\cal O}^u_{L(R)}U_{L(R)},\;\;
 D^\prime_{L(R)}={\cal O}^d_{L(R)}D_{L(R)},
\label{molr}
\end{equation}
with $U=(u,c,t)^T$ etc; the primed fields denote symmetry eigenstates and
the unprimed ones mass eigenstates, being the Cabibbo-Kobayashi-Maskawa
matrix  defined as $V_{\rm CKM}={\cal O}^{u T}_L{\cal O}^d_L$.

In terms of the mass eigenstates the Lagrangian interaction involving exotic
quarks, the known quarks, and doubly charged scalars is given by~\cite{cp3}:
\begin{equation}
-{\cal L}_Y=
\raisebox{4.5ex}{}
-\sqrt{2}\bar{J}\left[
e^{-i\theta_\chi}\frac{\vert v_\chi\vert}{N}\frac{M^d_\alpha}{\vert v_\rho\vert}
\,R-e^{+i\theta_\chi}\frac{\vert v_\rho\vert}{N} \frac{m_J}{\vert v_\chi\vert}
\,L
\right]({\cal O}^d_L )_{3\alpha}
d_{\alpha }Y^{++}
+ H.c.~,
\label{yuka}
\end{equation}
where $N$ is the same parameter appearing in Eq~(\ref{dcf}), i.e., 
$N =(\vert v^2_\rho\vert +\vert v^2_\chi\vert)^{1/2}$ and now, unlike 
Eq.~(\ref{yukawa}), all fields are mass 
eigenstates, $L=(1-\gamma_5)/2$, $R~=~(1~+~\gamma_5)/2$, with 
$m_J=\lambda_3\vert v_\chi\vert/\sqrt2$. In
writing  the first term of Eq.(\ref{yuka}) we have used $\tilde{F}_{3\alpha}=
\sqrt2({\cal O}^d_LM^d {\cal O}^{dT}_R)_{3\alpha}/\vert v_\rho\vert$,
where $M^d$ is the  diagonal mass matrix in the $d$-quark sector and we have
omitted the summation symbol in $\alpha$ so that $d_\alpha=d,s,b$. 
The Eq.~(\ref{yuka}) contains all CP violation in the quark sector once we 
have assumed that all the Yukawa couplings are real. Unlike in multi-Higgs extensions 
of the standard model~\cite{tdl,sw,wu,lavoura,desphande,sanda,donoghue,wolfen,chang} 
there is no Cabibbo suppression since in this model only one quark, $J$,
contributes in the internal line, i.e., we have the replacement $u,c,t,\to J$.

Notice that in Eq.~(\ref{yuka}) the suppression of the mixing angle in the
sector of the doubly charged scalars [see Eq.~(\ref{dcf})] has been
written explicitly. We will use as illustrative
values $\vert v_\rho\vert\leq246$ GeV and $\vert v_\chi\vert \stackrel{>}{\sim}
1$ TeV. In this situation the $CP$ violation in the neutral kaon system will
impose constraints only upon the masses $m_J$, $m_Y$, and, in principle, on
$m_U$ the mass of the doubly charged vector boson.
Although ${\cal O}^j_L$ has free parameters since the masses
$m_{j_{1,2}}$ are not known, the exotic quarks  $j_{1,2,}$
do not play any role in the CP violation phenomena of $K$ mesons.

We should mention that it was implicit in the model of Ref.~\cite{cp3} the
conservation of the quantum number B+L defined in Refs.~\cite{pt,mpp}. Only
in this circumstance (or by introducing appropriately a $Z_2$ symmetry) we can 
avoid terms like $\epsilon\overline{(\Psi_{aL})^c}\Psi_{bL}\eta$ and
$\overline{(l_{aL})^c}E_{bR}$, where $\Psi_L$,
$l_R$ and $E_R$ denote the left-handed lepton triplet, and the usual
right-handed components for usual and exotic leptons. These interactions imply
mixing among the left- and right-handed 
components of the usual charged leptons with the exotic ones~\cite{alex}. 
The quartic term $\chi^\dagger\eta\rho^\dagger\eta$ in the scalar potential
which would imply CP violation throughout the single charged scalar exchange is
also avoided by imposing the B+L conservation. In fact, this model has the
interesting feature that when a $Z_2$ symmetry is imposed, the Peccei-Quinn
$U(1)$, the total lepton number, and the barion number are all 
automatic symmetries of the classic Lagrangian~\cite{axion331}. 

\section{CP violation in the neutral kaons}
\label{sec:cpvkaons}

First of all let us say that in the present model there are tree level
contributions to the mass difference $\Delta M_K=2{\rm Re}M_{12}$ (where
$M_{12}=\langle K^0\vert{\cal H}_{eff} \vert \bar K^0\rangle/2m_K$).
This is because the existence of the flavor changing neutral
currents in the model in both the scalar sector and in the couplings with
the $Z^{\prime 0}$. The $H^0$'s contributions to $\Delta M_K$ have
been considered in Ref.~\cite{laplata1}. For $m_H\sim150$ GeV the
constraint coming from the experimental value of $\Delta M_K$ implies
$({\cal O}^d_L)_{dd}({\cal O}^d_L)_{ds}\lesssim0.01$.
There are also tree level contributions to
$\Delta M_K$ coming from the $Z^\prime$ exchange which were considered in
Ref.~\cite{331,laplata2}. 
However, since there are 520 diagrams contributing to
$\Delta M_K$, we will use in this work the experimental value
for this parameter. In this vain \textit{a priori} there is no constraints on
the matrix elements of ${\cal O}^d_L$. 

The definition for the relevant parameters in the neutral kaon system
is the usual one~\cite{gilman,gcb,bertolini,pdg}:
\begin{equation}
\epsl=\frac{e^{i (\delta_{2}-\delta_{0}+
\frac{\pi}{2})}}{\sqrt{2}}
\frac{{\rm Re} A_{2}}{{\rm Re}A_{0}}
\left[
\frac{{\rm Im} A_{2}}{{\rm Re} A_{2}}
-\frac{{\rm Im}A_{0}}{{\rm Re}A_{0}}
\right], \quad
\eps = \frac{e^{i\frac{\pi}{4}}}{\sqrt{2}}
\left[ \frac{{\rm Im} A_{0}}{{\rm Re} A_{0}}
+\frac{{\rm Im} M_{12}}{\Delta M_K}
\right],
\label{que}
\end{equation}
We shall use the $\Delta I=1/2$ rule for the nonleptonic decays which 
implies that ${\rm Re} A_{0} / {\rm Re} A_{2}\simeq 22.2$ and
that the phase $\delta_{2} - \delta_{0} \simeq -\frac{\pi}{4}$
is determined by hadronic parameters following Ref.~\cite{nir} and it is,
therefore, model independent.

The $\epsilon$ parameter has been extensively measured and its value is
reported to be~\cite{pdg}
\begin{equation}
\vert\epsilon_{exp}\vert=(2.284 \pm 0.014)\times 10^{-3}~.
\label{eexp}
\end{equation}
More recently, the experimental status for the
$\epsl / \eps$ ratio  has stressed the clear
evidence for a non-zero value and, therefore, the  existence of direct CP
violation. The present world average (wa) is~\cite{pdg}
\begin{equation}
\vert\epsilon^\prime / \epsilon\vert_{\textrm{wa}}=(1.67 \pm 0.26) \times 10^{-3}\; ,
\label{eepratio} 
\end{equation}
where the relative phase between $\eps$ and $\epsl$ is negligible~\cite{sozzi}.
These values of $\vert\eps\vert$ and $\vert\epsl/\eps\vert$ imply
\begin{equation}
\vert\epsilon^\prime_{exp}\vert=3.8\times 10^{-6}~.
\label{epexp}
\end{equation}

On the other hand, we can approximate
\begin{equation}
\vert\eps\vert
\approx \frac{1}{\sqrt{2}} \left\vert \frac{{\rm Im} M_{12}}{\Delta M_K  }\right\vert
~ \;(a),\quad
\vert\epsl\vert \approx   \frac{1}{22.2\sqrt{2}}\left\vert  
\frac{ {\rm Im} A_{0} }{ {\rm Re} A _{0} }\right\vert \;(b).
\label{epsl} 
\end{equation}

In the prediction of $\epsl / \eps$, ${\rm Re}A_{0}$ and $\Delta M_K$ are taken
from experiments, whereas ${\rm Im}  A_{0}$ and ${\rm Im} M_{12}$ are computed
quantities \cite{marco}. The experimental values used in this work are 
${\rm Re} A_{0}=3.3 \times 10^{-7}$ GeV and $\Delta M_K=3.5 \times 10^{-15}$ GeV.

Let us finally consider the condition with which we will calculate the
parameters $\epsilon$ and $\epsilon^\prime$. 
The main $\Delta S=1$ contribution for the  $\epsl$ parameter comes from the
gluonic penguin diagram in Fig.\,1 that exchanges a doubly charged scalar. 
The electroweak penguin is suppressed as in the SM
and will not be considered. On the other hand the $\Delta S=2$ and $CP$
violating parameter $\eps$ has only contributions coming from box diagrams
involving two doubly charged scalars $Y^{++}$ (see Fig.\,2a) and box diagrams
involving one doubly charged scalar and one vector boson $U^{++}$ (see
Fig.\,2b). The relevant vertices for the calculations are given in
Eq.~(\ref{yuka}) and we will use the unitary gauge in our calculations. In other
renormalizable $R_\xi$ gauges we must to take into account the would be 
Goldstone contributions and notice that, according to Eq.~(\ref{dcf}), the
component  of $\chi^{++}\sim O(1)G^{++}$.

The hadronic matrix elements will be taken from literature and  whenever
possible we also take, for the reasons we expose at the beginning of this
section, from the experimental data or as free parameters. 
One of the features of this model is that there is
no GIM  mechanism since the only CP violation source comes from the vertices
involving a $d$-type quark, an exotic quark, and a single doubly charged scalar.

\section{Direct CP violation}
\label{sec:epsilonprime}

The dominant contributions to the
$\epsilon^\prime$ parameter come from the penguin diagram showed in
the Fig.~1~\cite{bertolini,soviet}. The part of the Lagrangian  that takes into
account this amplitude is obtained from Eq.~(\ref{yuka}) and the corresponding
imaginary effective interaction is given by 
\begin{eqnarray}
\textrm{Im}\,{\cal L}_{\epsilon^\prime}=\frac{g_s}{16\pi^2N^2}\,
C_{ds}m_s
\left[ \bar{s}\,\sigma^{\mu\nu}\,\frac{\lambda_a}{2}
\left(L-\frac{m_d}{m_s}R\right)\,d\right]\,G^a_{\mu\nu}\,\frac{1}{2}\left[
h(x)-xh^\prime(x)\right]\sin 2\theta_\chi,
\label{p1}
\end{eqnarray}
where we have defined $C_{ds}=({\cal O}^d_L)_{3d}({\cal O}^d_L)_{3s}$, and
$G^a_{\mu\nu}$ in the
context of the effective interactions is just 
$G^a_{\mu\nu}=\partial_\mu G^a_\nu-\partial_\nu G^a_\mu$, $x=m^2_Y/m^2_J$ and 
the function $h(x)$ is given in the Appendix, and the prime denotes first 
derivative.

Neglecting the $\gamma,Z$ contributions, i. e., the amplitudes with $I=2$,
and using the values for the other parameters given above, 
Eq.~(\ref{epsl}b) leads to 
\begin{equation}
\vert\epsilon^\prime\vert\approx\frac{1}{\sqrt2}\frac{1}{22.2}
\frac{\vert\textrm{Im}A_0\vert}
 {\vert\textrm{Re}A_0\vert}
\approx 9.6\times10^4\; \frac{\vert\textrm{Im}\,A_0\vert}{1 \mathrm{GeV}},
\label{ep}
\end{equation}
where we have used ${\rm Re}A_0=3.3 \times 10^{-7}\textrm{GeV}^{-1}$, with,
\begin{equation}
\vert {\rm Im}\,A_0\vert= \sqrt{3}\frac{g_s}{16\pi^2}\frac{m_s}{N^2}\,C_{ds} 
\left\vert \frac{1}{2}\left( h(x)-xh^\prime(x)\right)\right\vert \left\vert
P_L-\frac{m_d}{m_s}P_R\right\vert
\sin2\theta_\chi.
\label{ep2}
\end{equation}
We can write $\vert\epsilon^\prime\vert$ as follows:
\begin{equation}
\frac{\vert \epsilon^\prime\vert}{\vert\epsilon^\prime_{exp}\vert}
=C_{ds}A(x)\sin2\theta_\chi,
\label{new}
\end{equation}

\begin{equation}
A(x)=\sqrt3\frac{g_s}{(4\pi)^2}
\frac{m_s}{\vert\epsilon^\prime_{exp}\vert N^2}
\left\vert P_L-\frac{m_s}{m_d}P_R\right\vert
\left\vert \frac{1}{2}\left(h(x)-xh^\prime(x)\right)\right\vert \;
\frac{9.6\times10^4}{1\textrm{GeV}},
\label{adef}
\end{equation}
where we have defined the matrix elements
\begin{equation}
P_L=\langle\pi \pi (I=0)\vert
 \;(\bar{s}
\sigma^{\mu \nu} L \frac{\lambda^a}{2} d) \;
G^a_{\mu\nu} \;\vert K^0 \rangle,
\quad
P_R=\langle\pi \pi (I=0)\vert \;(\bar{s}
\sigma^{\mu \nu} R \frac{\lambda^a}{2} d)\;
G^a_{\mu\nu} \;\vert K^0 \rangle.
\label{brl}
\end{equation}

Using the bag model (BM) it has been obtained that
$P_L=-0.5\,\textrm{GeV}^2$~\cite{donoghue}. The other term in Eq.~(\ref{ep2})
with the matrix element $P_R$ is negligible [even if $\vert P_R\vert\approx O(
\vert P_L\vert)$] since it has a $m_d$ factor. We will also use the following 
values: $m_K=498\;{\rm MeV}$,  $m_d/m_s=1/20$,
$m_s=120\;{\rm MeV}$, and $\alpha_s=0.2$. The function $\vert h(x)-
xh^\prime(x)\vert$ has its maximum equal to one at $x=0$. Both $P_L$ and $P_R$
matrix elements can be considered as free parameters, for instance in 
Fig.~\ref{fig3} we use $P_L=(1/2)P_L(BM)$. Of course, there is also a 
solution if we use the bag model value of $P_L$.

\begin{figure}
\begin{picture}(260,180)(20,-20)
\SetWidth{0.5}
\DashCArc(150,91)(60,0,180){8}
\ArrowLine(45,91)(90,91)
\ArrowLine(210,91)(255,91)
\Gluon(150,91)(150,-14){7.5}{6}
\Text(30,93)[cc]{{$d$}}
\Text(270,92)[cc]{{$\bar{s}$}}
\Text(180,76)[cb]{{$J$}}
\Text(120,76)[cb]{{$J$}}
\SetWidth{0.5}
\ArrowLine(90,91)(150,91)
\ArrowLine(150,91)(210,91)
\Text(135,136)[lb]{{$Y^{++}$}}
\Text(165,31)[lb]{{$g$}}
\end{picture}
\label{fig1}
\caption{Dominant CP violating penguin diagram contributing to the decay
$K^0\rightarrow \pi\pi$.}
\end{figure}
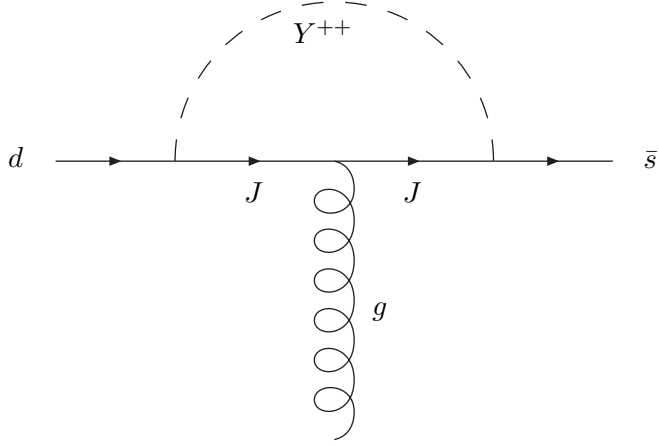

\section{Indirect CP violation}
\label{sec:epsilon}

The contributing diagrams for the $\eps$ parameter are of two types,
one with the exchange of two $Y^{++}$ and the other with one $U^{++}$ and
one $Y^{++}$. They are shown in the Figs. 2a and 2b, respectively. The
imaginary part for this class of diagrams has been derived in
Refs.~\cite{wolfen} and \cite{chang}. The  Higgs scalar-quark 
interaction is  given in Eq.(\ref{yuka}) and the 
gauge boson-quark Lagrangian interaction is 
\begin{equation}
{\cal L}_W=-\frac{g}{\sqrt{2}} \bar{J} ({\cal O}^d_L)_{3 \alpha}
\gamma^{\mu} L d_{\alpha} U^{++}_{\mu} + H.c.
\label{droga}
\end{equation}

The contributions to the effective Lagrangian of diagrams like that shown in 
Fig. 2a are given by
\begin{eqnarray}
\textrm{Im}{\cal L}^{YY}_\epsilon&=&\frac{C^2_{ds}}{(4\pi)^2}\frac{2m^2_K}{N^2}
\frac{m^2_s}{N^2}\left\{\left[\frac{\sin4\theta_\chi}{m^2_K}[(\bar{s}Ld)^2-
\frac{m^2_d}{m^2_s}(\bar{s}Rd)^2]\right]g_0(x)\right.\nonumber \\ &-&\left.
\frac{\vert v_\rho\vert}{4\vert
v_\chi\vert}\sin2\theta_\chi\left(\left[\bar{s}\gamma^\mu Li\stackrel{\leftrightarrow}
{\partial}_\mu d\bar{s}\left(L-\frac{m_d}{m_s}R\right)d\right]
\left[5g_0(x)+\frac{3}{2}xg^\prime_0(x) \right]\right.\right.
\nonumber \\ &+& \left.\left.
\bar{s}\left(L-\frac{m_d}{m_s}R\right)i\stackrel{\leftrightarrow}
{\partial}_\mu d\bar{s}\gamma^\mu Ld\left[g_0(x)+\frac{3}{2}xg^\prime_0(x)
\right]\right)\right\},
\label{box1}
\end{eqnarray}
where $g_0(x)$ is given in the Appendix. 

On the other hand, the contributions to the effective Lagrangian of diagrams like
that shown in Fig.~2.b are given by 
\begin{eqnarray}
\textrm{Im}{\cal L}^{UY}_\epsilon&=&\frac{C^2_{ds}}{(4\pi)^2}
\frac{2m^2_K}{N^2}\frac{m^2_s}{N^2}\left(\frac{g^2}{2}\frac{N^2}{4m^2_J}\right)
\frac{\sin2\theta_\chi}{m_sm^2_K}\left\{\left[ 
\bar{s}\gamma^\mu\gamma^\nu\left(L-\frac{m_d}{m_s}R\right)i
\stackrel{\leftrightarrow}{\partial}_\mu d\right](\bar{s}\gamma_\nu L d)\,
\,E_1(x,y)\right.\nonumber \\&+&
\left.
\bar{s}\gamma_\nu Li\stackrel{\leftrightarrow}{\partial}_\mu  d
\bar{s}\gamma^\mu\gamma^\nu\left(L-\frac{m_d}{m_s}R\right)d \, E_2(x,y) 
+\left[\bar{s}\left(L-\frac{m_d}{m_s}R\right)
i\stackrel{\leftrightarrow}{\partial}_\mu  d (\bar{s}\gamma^\mu
Ld)\right.\right.\nonumber \\&+&\left.\left.\bar{s}\gamma^\mu L 
i\stackrel{\leftrightarrow}{\partial}_\mu d
\bar{s}\left(L-\frac{m_d}{m_s}R\right)d\right]\,E_3(x,y)
-\left[i\partial_\mu\left[\bar{s}\gamma^\mu\gamma^\nu
\left(L-\frac{m_d}{m_s}R\right)d\right] \bar{s}\gamma_\nu Ld
\right.\right.\nonumber \\ &+&\left.\left.
i\partial_\mu ( \bar{s}\gamma_\nu L d)
\bar{s}\gamma^\mu\gamma^\nu\left(L-\frac{m_d}{m_s}\right)d\right]\,E_4(x,y)-
\left( i\partial_\mu
\left[\bar{s}\left(L-\frac{m_d}{m_s}R\right)d\right]\bar{s}\gamma^\mu Ld \
\right.\right.\nonumber \\ &+&\left.\left.
i\partial_\mu (\bar{s}\gamma^\mu Ld)\bar{s}
\left(L-\frac{m_d}{m_s}R\right)d\right)\,E_5(x,y)
\right\},
\label{box2}
\end{eqnarray}
where $y=m^2_U/m^2_J$ and the functions $E_{1,2,3,4,5}$ are defined in the 
Appendix. 

Taking into account both contributions in Eqs.(\ref{box1}) and (\ref{box2}) and
using  
\begin{equation}
\textrm{Im}\,M_{12}=\frac{\textrm{Im}\langle 
\bar{K}^0 \vert {\cal L}_\epsilon(0)\vert K^0\rangle}{2m_K},
\label{m12}
\end{equation}
we obtain

\begin{eqnarray}
\textrm{Im}\,M_{12}&=&-\frac{C^2_{ds}}{(4\pi)^2}\frac{m^2_K}{N^2}\frac{f^2_K}
{2N^2}\left(1+\frac{m_d}{m_s}\right)^{-2}
\left\{
\frac{5}{6}
\sin4\theta_\chi\left(1-\frac{m^2_d}{m^2_s} \right)g_0(x)
 \right.\nonumber \\&-&\left.
 \frac{\vert v_\rho\vert}{2\vert v_\chi\vert}\sin2\theta_\chi
 \left[
 \frac{5}{12}
 \left( 5g_0(x)-\frac{3}{2}g^\prime_0(x)\right)
 \right.\right. \nonumber \\&-&\left.\left.
 \frac{1}{3}\left[ 1+\frac{1}{4}
 \left(
\frac{m_s+m_d}{m_K}
\right)^2
\right]
\left(g_0(x)+\frac{3}{2}xg^\prime_0(x)\right)
 \right]
 \right.\nonumber \\&+&\left.
 \frac{g^2}{2}\frac{N^2}{2m^2_J}\sin2\theta_\chi
 \left[
 \frac{2}{3}
 \left( E^\prime_1(x,y)+E_3(x,y)\right) -\frac{2}{3}\left( \frac{m_s+m_d}{m_K}
 \right)^2\left[ E_1(x,y)+E_4(x,y)\right]
 \right.\right.
 \nonumber \\&+& \left.\left. \frac{1}{12}\left[ 1-\left( \frac{m_s+m_d}{m_K}
 \right)^2\right]\left( E_2(x,y)+E_4(x,y)\right)\right]
\right\}.
\label{imm}
\end{eqnarray}

Thus, we can calculate $\vert\epsilon\vert$ from Eq.~(\ref{epsl}a) using 
$f_K=161.8$ MeV and $\Delta M_K=3.5\times10^{-15}$ GeV~\cite{pdg}. 
We have used the vacuum insertion (VI) approximation, and obtained: 
\begin{eqnarray}
B_L=\langle \bar{K}^0\,\vert[\bar{s}L(R)d]^2\vert
K^0\rangle=-\frac{5}{12}\,\frac{m^4_Kf^2_K}{(m_s+m_d)^2},\nonumber \\
\langle \bar{K}^0\vert [\bar{s}\gamma^\mu R(L)d]^2\vert K^0\rangle\simeq
\frac{2}{3}\,f^2_Km^2_K\,\nonumber \\
\langle\bar{K}^0\vert\bar{s} Rd\bar{s}Ld\vert K^0\rangle=
\frac{1}{2}\frac{m^4_Kf^2_K}{(m_s+m_d)^2}+\frac{1}{12}m^2_Kf^2_K,
\nonumber \\
\langle\bar{K}^0\vert \bar{s}\gamma^\mu Ld\bar{s}L(i\partial_\mu)d 
\vert K^0\rangle=-\frac{5}{6}\frac{m_dm^4_Kf^2_K}{(m_s+m_d)^2},\nonumber \\
\langle \bar{K}^0\vert \bar{s}\gamma^\mu Ld(-i\partial_\mu)\bar{s} Ld\vert
K^0\rangle=\frac{1}{3}\frac{m_sm^4_Kf^2_K}{(m_s+m_d)^2}+\frac{1}{12}m_sm^2_K
f^2_K, \nonumber \\
\langle \bar{K}^0\vert (-i\partial_\mu)\bar{s}\gamma_\nu
Ld\bar{s}\gamma^mu\gamma^\nu Ld\vert K^0\rangle=-\frac{2}{3}
\frac{m_sm^4_Kf^2_K}{(m_s+m_d)^2}.
\label{opevi}
\end{eqnarray}

We have verified that the main contribution to the box diagrams in 
Eqs.~(\ref{box1}) comes from the matrix element denoted 
by $B_L$. Thus, in order not to be restricted to the VI 
approximation, $B_L$ can be considered a free parameter and, for instance
in Fig.~\ref{fig3}, we have used $B_L=3B_L(VI)$, but there is also a solution 
using the VI value of $B_L$. 
\begin{center}
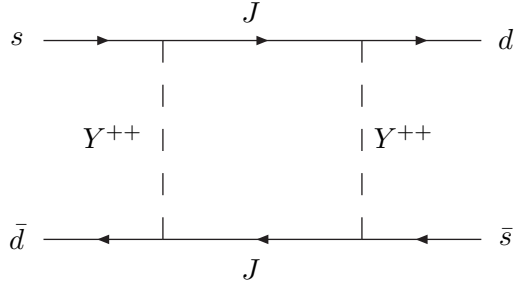
\begin{figure}
\begin{picture}(230,117)(15,-20)
\SetWidth{0.5}
\ArrowLine(165,1)(90,1)
\ArrowLine(210,1)(165,1)
\ArrowLine(45,76)(90,76)
\ArrowLine(90,1)(45,1)
\DashLine(90,76)(90,1){8}
\DashLine(165,1)(165,76){8}
\ArrowLine(90,76)(165,76)
\ArrowLine(165,76)(210,76)
\Text(35,77)[cc]{{$s$}}
\Text(35,2)[cc]{{$\bar{d}$}}
\Text(220,77)[cc]{{$d$}}
\Text(220,2)[cc]{{$\bar{s}$}}
\Text(60,40)[lc]{{$Y^{++}$}}
\Text(170,40)[lc]{{$Y^{++}$}}
\Text(120,91)[lt]{{$J$}}
\Text(120,-14)[lb]{{$J$}}
\end{picture}
\renewcommand{\thefigure}{2a}
\caption{\label{fig2a}
One of the box diagrams responsible for the transition $\bar{K}^0\rightarrow
K^0$ that involves the exchange of two doubly charged scalars $Y^{++}$.}
\end{figure}
\end{center}
\begin{center}
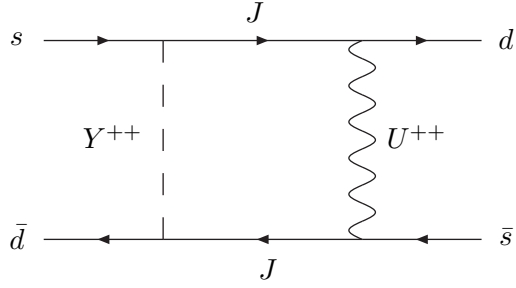
\begin{figure}
\begin{picture}(230,117)(15,-20)
\SetWidth{0.5}
\ArrowLine(165,1)(90,1)
\ArrowLine(210,1)(165,1)
\ArrowLine(45,76)(90,76)
\ArrowLine(90,1)(45,1)
\ArrowLine(90,76)(165,76)
\ArrowLine(165,76)(210,76)
\Text(35,77)[cc]{{$s$}}
\Text(35,2)[cc]{{$\bar{d}$}}
\Text(220,77)[cc]{{$d$}}
\Text(220,2)[cc]{{$\bar{s}$}}
\Text(60,40)[lc]{{$Y^{++}$}}
\Text(175,40)[lc]{{$U^{++}$}}
\Text(125,91)[ct]{{$J$}}
\Text(130,-14)[cb]{{$J$}}
\SetWidth{0.5}
\DashLine(90,76)(90,1){8}
\Photon(165,1)(165,76){5}{5.5}
\end{picture}
\renewcommand{\thefigure}{2b}
\addtocounter{figure}{-1}
\caption{\label{fig2b}
One of the box diagrams responsible for the transition $\bar{K}^0\rightarrow
K^0$ that involves the exchange of a doubly charged scalar $Y^{++}$ and a
doubly charged vector boson $U^{++}$.}
\end{figure}
\end{center}

\section{Fitting the experimental values}
\label{sec:fitting}

In order to compare the prediction of the model with the experimental data for
the CP violation in the neutral kaon system we use Eqs.~(\ref{ep})--(\ref{brl})
for $\vert\epsilon^\prime\vert$ and rewrite Eq.~(\ref{imm}) for 
$\vert \epsilon\vert$ as
\begin{equation}
\frac{\vert\epsilon\vert}{\vert \epsilon_{exp}\vert}=
C^2_{ds}B(x)\left(\frac{1}{2}\sin4\theta_\chi-b(x,y)
\sin2\theta_\chi\right).
\label{eteo2}
\end{equation}

where 

\begin{eqnarray}
B(x)&=&
\frac{1}{(4\pi)^2\vert\epsilon_{exp}\vert}
\frac{m^2_K}{N^2}\frac{f^2_K}{2N^2}
\frac{\sqrt{2}m_K}{\Delta M_K}
\big(1+\frac{m_d}{m_s}\big)^{-2}
\big(1-\frac{m^2_d}{m^2_s}\big)
\frac{5}{6}g_0(x)
\cr
&\approx&
1.34\times
\bigg(\frac{1\mathrm{TeV}}{N}\bigg)^4
g_0(x)
\label{bdef}
\end{eqnarray}

\begin{eqnarray}
b(x,y)&=&
\frac{6}{5}\big(1-\frac{m^2_d}{m^2_s}\big)^{-1}\;
\frac{1}{g_0(x)}
\Bigg\{
\frac{\vert v_\rho\vert}{4\vert v_\chi\vert}
\bigg\{
\frac{5}{12}\left[5g_0(x)-\frac{3}{2}xg'_0(x) \right] 
\nonumber \\ &-&
\frac{1}{3}\left[1+\frac{1}{4}\left(\frac{m_s+m_d}{m_K}\right)^2\right]
\left[g_0(x)+\frac{3}{2}xg'_0(x)\right]\bigg\} \nonumber \\ &-&
\frac{g^2}{2}\frac{N^2}{4m^2_J} 
\bigg\{
\frac{2}{3}\left[E_2(x,y)+E_4(x,y)\right]
-\frac{2}{3}\left(\frac{m_s+m_d}{m_K}\right)^2\left[E_1(x,y)+E_4(x,y)\right]
\nonumber \\ &+&\frac{1}{12}\left[1-\left(\frac{m_s+m_d}{m_K}\right)^2\right]
\big(E_3(x,y)+E_5(x,y)\big)
\bigg\}
\Bigg\}.
\label{bdef2}
\end{eqnarray}

Next we use the constraints
\begin{equation}
\left\vert\frac{\epsilon^\prime(C_{ds},x,\theta_\chi)}
{\epsilon^\prime_{exp}}\right\vert
=1,\quad \left\vert\frac{\epsilon(C_{ds},x,y,\theta_\chi)}
{\epsilon_{exp}}\right\vert=1.
\label{fit1}
\end{equation}
Notice that the above conditions are the strongest since we are not considering
the experimental error.

After some algebraic manipulations the constraints in Eqs.~(\ref{fit1}) imply
\begin{equation}
C^2_{ds}=\frac{D^4(x)}{
D^2(x)-\frac{b^2(x,y)}{A^2(x)}-
\frac{b(x,y)A(x)}{B(x)} 
}\leq1,
\label{cds}
\end{equation}
and
\begin{equation}
C_{ds}\sin\theta_\chi=\frac{1}{A(x)},
\label{sintheta}
\end{equation}
where we have defined
\begin{equation}
D^2(x)=\frac{1}{A^2(x)}+\frac{A^2(x)}{B^2(x)},
\label{ddef}
\end{equation}
with $A(x)$ defined in Eq.~(\ref{adef}), and $B(x)$ and $b(x,y)$ 
were defined in Eqs.~(\ref{bdef}) and (\ref{bdef2}), respectively.

It is interesting to note that
\begin{equation}
C_{ds}\sin2\theta_\chi \geq
\frac{1}{A(0)}=0.072\;\left(\frac{N}{1\,{\rm TeV}}\right)^2,
\label{f1}
\end{equation}
where we have used the value of the parameters as discussed below Eq.~(\ref{brl}).

We have study numerically Eq.~(\ref{cds}) and (\ref{sintheta}) and verified that they
are sensible to the values of the matrix elements $P_L$ in Eq.~(\ref{brl}) and $B_L$ 
defined in Eq.~(\ref{opevi}).



\begin{figure}[ptb]
\begin{center}
\includegraphics[width=8.5cm,height=12cm,angle=-90]{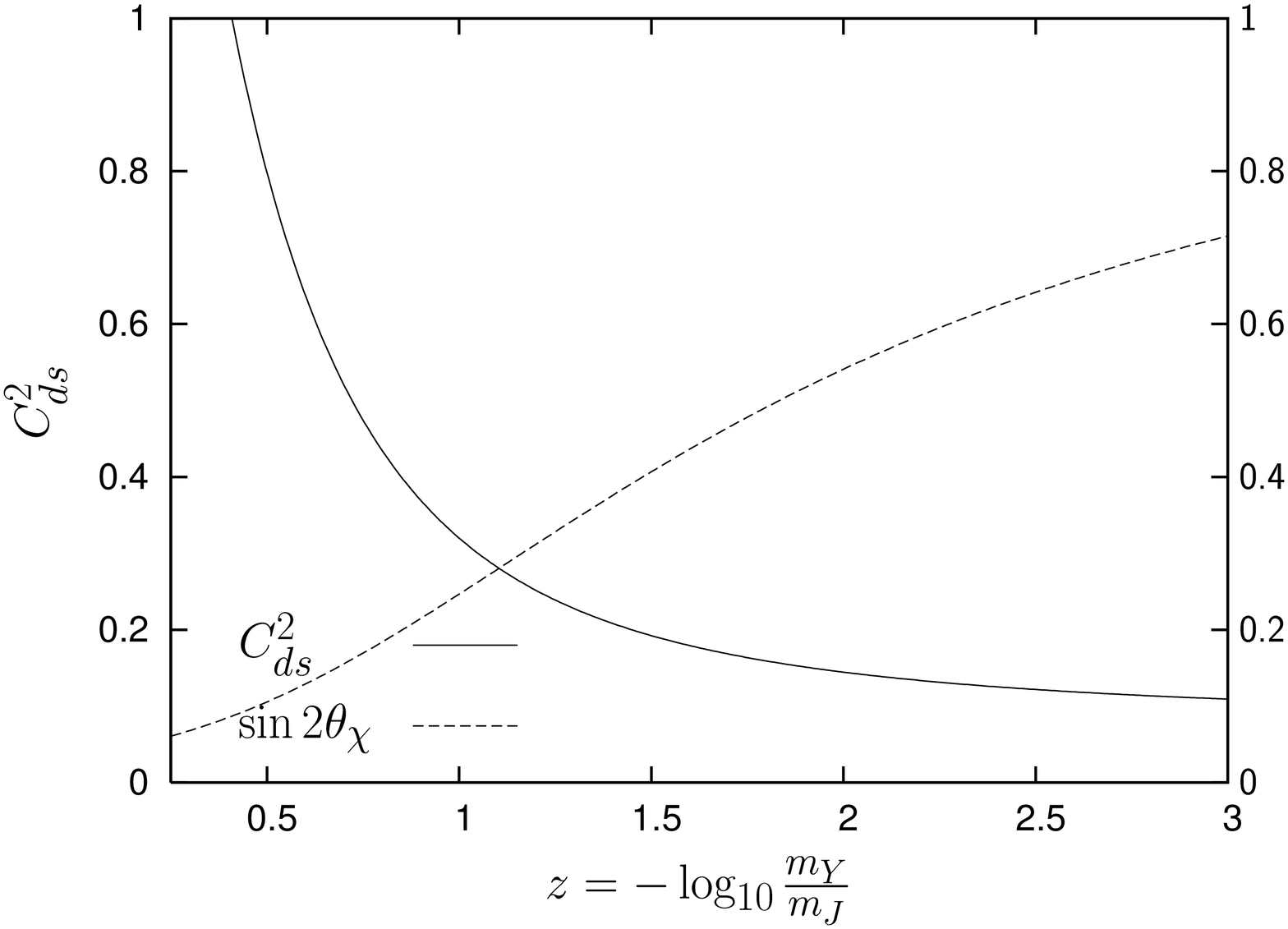}
\end{center}
\caption{\label{fig3}
Using Eq.~(\ref{cds}) and Eq.~(\ref{sintheta}) we studied the $x$-dependence of 
$C_{ds}^2$ (left scale) and $\sin 2\theta_{\chi}$ (right
scale) on $z$, respectively, with $z$ defined by $10^{-z}=m_Y/m_J=\sqrt{x}$. 
We have used $P_L=(1/2)P_L(BM)$ and $B_L\to 3B_L(VI)$, where BM indicates the
value of $P_L$ in the bag model, and VI means the  
vacuum insertion value of $B_L$.
We have also used $N=0.7$ TeV and $m_U/m_J=1$.
Notice that $\sin2\theta_{\chi}$ does not depend on $N$.
}
\end{figure}

The curves in Fig.~\ref{fig3} are curves of compatibility with experimental
data according to the constraints in Eq.~(\ref{fit1}). The dashed curve 
shows all the allowed value for $\sin2\theta_\chi$ while the continue curve
shows the allowed values for $C_{ds}$ as a function of $x$. However, these 
values are not independent from one another if we want to satisfy both
constraints at the same time. The compatibility with the experimental data
is obtained by drawing a vertical line for a given value of $z$. For instance
using $z=2$ (i. e., $m_J=100\,m_Y$) we found $\sin2\theta_\chi\approx 0.15$ and
$C^2_{ds}\approx 0.6$, for $z=1$ we obtain $\sin2\theta_\chi\approx0.25$ and
$C^2_{ds}\approx 0.3$. Notice that from Fig.~\ref{fig3} we  see that we have
solution in the range $2.5^o\lesssim \theta_\chi\lesssim 22.5^o$.

\section{Conclusions and discussions}
\label{sec:con}

The study showed that the 3-3-1 model considered here can account for
the direct and indirect CP violation present in the $K^0-\bar{K}^0$ system 
for sensible values of the unknown parameters. 
Within the approximations used, 
$N\lesssim 1{\rm TeV}$ there are infinitely many possible values for
$C_{ds}$, $\theta_{\chi}$ and $m^2_Y/m^2_J$ allowed by the
experimental data. Although they are not all
independent and the constraint $\vert C_{ds}\vert<1$ implies a very small 
upper bound for the ratio $m^2_Y/m^2_J$. Such bound becomes smaller as
$N$ becomes greater. Thus very large values of $N$ leads to
unrealistically small values of the ratio. Notice also that the constraints 
used in Eq.~(\ref{fit1}) are very strong. However, weaker constraints arise 
if a detailed analysis which take into account the experimental error in 
both $\epsilon$ and $\epsilon^\prime$ is done. Of course, it is clear that
in this case there will exist a solution as well.

The model implies also some contributions to the neutron electric 
dipole moment (EDM) as in Ref.~\cite{cp3}
\begin{equation}
d_n\simeq 4.9\times10^{-22}\,\left[\sum_\alpha G_{\alpha 1}
({\cal O}^u_R)_{1\alpha}
({\cal O}^u_L)_{11}\,\sin\theta_\chi\right]\;\textrm{e}\,\textrm{cm}, 
\label{edmn}
\end{equation}
and we see that a value compatible with the experimental bound 
of~\cite{harris} 
\begin{equation}
\vert d_n\vert< 6.3\times10^{-26}\;\textrm{e}\;\textrm{cm}\;\;
(90 \textrm{\% CL}),
\label{nedm}
\end{equation} 
is obtained for practically any value of the phase $\theta_\chi$, if 
$ \sum_\alpha G_{\alpha 1}({\cal O}^u_R)_{1\alpha}
({\cal O}^u_L)_{11}\sim 10^{-5}$.
The EDM of the charged leptons also produces results compatible with the
experimental limit for a large range of the parameters of the model. 
In addition this model allow magnetic dipole moments for massive neutrinos
in the range $10^{-13}-10^{-11}\,\mu_B$ almost independently of the neutrino 
mass~\cite{mp99} which is near the experimental upper limit for the 
electron neutrino magnetic moment~\cite{munu} 
\begin{equation}
\mu_e<10^{-11}\;\mu_B\;\; (90 \textrm{\% CL}).
\label{mnue}
\end{equation}

Moreover, as in the standard model the lepton charge asymmetry 
in the $K_{l3}$ decay, $\delta_L$, which has the experimental 
value (the weighted average of $\delta(\mu)$ and $\delta(e)$~\cite{pdg}) 
$\delta_L~=~(3.27\pm~0.12)~\times 10^{-3}$, is also automatically fitted in 
the present model because $\vert A(K^0\to\pi^-e^+\nu_e)\vert= 
\vert A(\bar{K}^0~\to~\pi^+e^-\bar{\nu}_e)\vert$ is still valid. 

Recent analysis on CP violation indicate that the
phase of the CKM matrix, which is $O(1)$, is the dominant contribution to the
CP violation in both $K$ and $B$ mesons so, new phases
coming from physics beyond the standard model must be small perturbations.
The CKM mechanism is also at work in the present model but we switch it off 
in order to study the possibilities of the extra phase of the model. 
Concerning the $K$ meson and EDM for elementary particles it seems that
the model do well. Presently we are working out the case of $B$ decays,
if the model is not able to fit these data it implies that CKM phase
must be switched on. It is still possible that new phases
may be at work if decays based on $b\to s$ gluonic dominated
transition really need new physics~\cite{soni05}. 
Any way, the extra phase in the model could be
important for other CP violation parameters like the EDM or, if new
CP violation observables in $B$-mesons will not be fitted by the CMK 
mechanism.

Finally, some remarks concerning the masses of the extra particles in 3-3-1 
models. Firstly, let us consider the $Z^\prime$ vector boson it
contributes to the $\Delta M_K$ at the tree level so that there is 
a constraint over the quantity~\cite{dumm94,sher}
\begin{equation}
({\cal O}^d_L)_{3d}({\cal O}^d_L)_{3s}\,
\frac{M_Z}{M_{Z^\prime}},
\label{ufa}
\end{equation}
which must be of the order of $10^{-4}$ to have compatibility with the
measured $\Delta M_K$. This can be achieved with $M_{Z^\prime}\sim 4$ TeV 
if we assume a Fritzsch-structure ${\cal O}^d_{Lij}=\sqrt{m_j/m_i}$ or, 
since there is no \textit{a priori} reason for ${\cal O}^d_L$ having
the Fritzsch-structure, it is possible that the product of the mixing
angles saturates the value $10^{-4}$~\cite{dumm94}, in this case $Z^\prime$ 
can have a mass near the electroweak scale. However, in 3-3-1 models 
there are flavor changing neutral currents  in the scalar sector implying 
new contributions to $\Delta M_K$ which are of the form 
\begin{equation}
({\cal O}^d_L)_{d3}\Gamma^d_{3\beta}({\cal O}_R)_{\beta s}\,\frac{M_Z}{M_H},
\label{scalarmk}
\end{equation} 
that involve the mass of the scalar $M_H$, the unknown matrix elements
${\cal O}^d_R$ and also the Yukawa coupling $\Gamma^d$, so their
contributions to $\Delta M_K$ can have opposite sing relative to that 
of the $Z^\prime$ contribution. This calculation has not been done in 
literature, where only the later contribution has been taken into 
account~\cite{dumm94,sher}.
The model has also doubly charged scalars that are important in the
present CP violating mechanism. The lower limit for the
mass of doubly charged scalars is a little bit above 100 GeV~\cite{d0}.
Concerning the doubly charged vector boson, if they have masses above 500 
GeV they can be found (if they really do exist) by measuring left-right
asymmetries in lepton-lepton scattering~\cite{marcos}.
Fermion pair production at LEP, and lepton flavor violating of the charged
leptons suggest a low bound of 750 GeV for the $U^{--}$ mass~\cite{tully}.
In $e^+e^-,e\gamma$, $\gamma\gamma$ colliders the 
detection of bileptons with  masses between 500 GeV and 1 
TeV~\cite{dion1} is favored, while if their masses are of the 
order of  $\stackrel{<}{\sim}$~1~TeV they could be also observed 
at hadron colliders like LHC \cite{dion2}.
Muonium-antimuonium transitions would imply a lower bound of 850 GeV on
the masses of the doubly charged gauge bileptons, 
$U^{--}$~\cite{willmann99}. However this bound depends on
assumptions on the mixing matrix in the lepton charged currents coupled 
to $U^{--}$ and also it does not take into account that there are in the model 
doubly charged scalar bileptons which also contribute to that 
transition~\cite{pleitez00}.
The muonium fine structure only implies $m_U/g>215$ GeV assuming only
the vector bilepton contributions~\cite{fujii}. 
Concerning the exotic quark masses, there is no lower limit for their 
masses but if they are in the range of 200-600 GeV they may be 
discovered at the LHC~\cite{yara}.
Search for free stable color triplets quarks 
has been carried out in 
$p\bar{p}$ collider at an energy of 1.8 GeV excluding these particles 
in the range 50-139 GeV, 50-116 GeV and 50-140 GeV, for the electric charges
of $+1$, $2/3$ and $4/3$, respectively~\cite{abe}. We can conclude that the
masses for the extra degrees of freedom which distinguish 3-3-1 models
with respect to the standard model may be accessible at the energies of the 
colliders of the next generations. 

\acknowledgments
This work was partially supported by CNPq under the processes
141874/03-1 (CCN), 305185/03-9 (JCM) and 306087/88-0 (VP) and by FAPESP 
(MCR and OR).

\appendix

\section{Integrals}
\label{app:inte}

\begin{equation}
g_0(x,y)=-\frac{1}{x-y}\left[\left(\frac{x}{x-1}\right)^2\ln x-
\left(\frac{y}{y-1}\right)^2\ln y-\frac{1}{x-1}+\frac{1}{y-1}\right]
\label{g0}
\end{equation}

\begin{equation}
g_1(x,y)=-\frac{1}{x-y}\left[\frac{x}{(x-1)^2}\ln x-
\frac{y}{(y-1)^2}\ln y-\frac{1}{x-1}+\frac{1}{y-1}\right]
\label{g1}
\end{equation}

\begin{equation}
g_0(x)=\lim_{y\to x} g_0(x,y)=-\frac{2}{(x-1)^2}+\frac{x+1}{(x-1)^3} \ln x
\label{g0x}
\end{equation}

\begin{equation}
h(x)=-\frac{x}{(x-1)^2}\ln x+\frac{1}{x-1},\quad h^\prime(x)=
\frac{2-2x+(1+x)\ln x}{(x-1)^3},
\label{hx}
\end{equation}

\begin{equation}
E_1(x,y)=\left(\frac{5}{2}+\frac{1}{4}(x\partial_x+y\partial_y)
\right)g_0(x,y)-\frac{5}{8y}g_1(x,y)
\label{e1}
\end{equation}

\begin{equation}
E_2(x,y)=\left(\frac{1}{2}+\frac{1}{4}(x\partial_x+y\partial_y)
\right)g_0(x,y)-\frac{1}{8y}g_1(x,y)
\label{e1p}
\end{equation}

\begin{equation}
E_3(x,y)=\frac{3}{4y}\left(1+x\partial_x+y\partial_y\right)g_1(x,y).
\label{e2}
\end{equation}

\begin{equation}
E_4(x,y)=\frac{1}{4}\left[(x\partial_x-y\partial_y)g_0(x,y)+
\frac{1}{2y}g_1(x,y)\right]
\label{e3}
\end{equation}

\begin{equation}
E_5(x,y)=-\frac{1}{4y}\left(3+x\partial_x-y\partial_y\right)g_1(x,y)
\label{e4}
\end{equation}


\end{document}